\documentclass[aps,10pt,prb,twocolumn,superscriptaddress,longbibliography,floatfix]{revtex4-2}
\usepackage{diagbox} 
\usepackage{graphicx}
\usepackage{pstricks}
\usepackage{pst-node}
\usepackage{pstricks-add}
\usepackage{tikz}
\usepackage{amsfonts,amssymb,amsmath}
\usepackage[colorlinks=true,citecolor=green,linkcolor=red,urlcolor=blue]{hyperref}
\usepackage{subfigure}
\usepackage{calligra}
\usepackage{braket}
\usepackage{dsfont}

\usepackage{soul}

\newcommand{\beq}{\begin{equation}}
\newcommand{\be}{\begin{equation}}
\newcommand{\beqn}{\begin{eqnarray}}
\newcommand{\eeq}{\end{equation}}
\newcommand{\ee}{\end{equation}}
\newcommand{\eeqn}{\end{eqnarray}}

\begin{document}
\title{Fermionic many-body topology in cavity-coupled Su-Schrieffer-Heeger chains}
\author{Anna Ritz-Zwilling}
\affiliation{CPHT, CNRS, École polytechnique, Institut Polytechnique de Paris, Palaiseau, France}

\author{Olesia Dmytruk}
\affiliation{CPHT, CNRS, École polytechnique, Institut Polytechnique de Paris, Palaiseau, France}


\begin{abstract}

We study the effects of cavity-mediated interactions on the symmetries and topological properties of a Su-Schrieffer-Heeger chain coupled to a single mode photonic cavity. Considering an off-resonant regime we use the high-frequency expansion in order to obtain an effective fermionic Hamiltonian with cavity-mediated interactions. We characterize the effects of the cavity on topology in a finite size chain by studying three different fermionic many-body topological markers using exact diagonalization: correlation functions between edges in a chain with open boundary conditions, and a winding number based on the single-particle Green's function and bulk electric polarization via the many-body formula by Resta for a chain with periodic boundary conditions. 
Importantly, we show that chiral symmetry is not broken by the cavity in the many-body ground state in the high-frequency limit. Our work highlights the importance of the cavity-mediated many-body interactions in understanding the symmetry and topology of the system.

\end{abstract}

\date{\today}

\maketitle

\section{Introduction}

Coupling electronic systems with cavity photons realizes new means to investigate and manipulate properties of materials  through light–matter interactions \cite{Schlawin22, GarciaVidal21}.  For instance, modification of quantum Hall effect~\cite{appugliese2022breakdown,enkner2025}, control of metal-to-insulator transition~\cite{jarcNature2023}, and cavity-altered superconductivity~\cite{Keren2026cavity} were experimentally achieved with cavity embedding. On the theory side, effect of cavity coupling on quantum Hall systems~\cite{ciuti2021cavity,winter2025fractional,borici2025}, superconductors\cite{sentef2018,schlawin2019cavity,curtis2019cavity,alloca2019,kozin2025cavity}, correlated matter~\cite{mazza2019,passetti2023cavity,kass2024manybody}, and moir\'e  materials~\cite{nguyen2023electron} was studied in recent years. The cavity control of topological properties of materials, such as topological superconductors hosting Majorana bound states~\cite{trif2012resonantly,cottet2013squeezing,mendez2020renyi,contamin2021topological,Bacciconi24,dmytruk2024hybrid,fernandez25,kobialka2026topology} and topological insulators~\cite{Dmytruk22,Perez25,vlasiuk2023cavity,Ciuti24,Shaffer24,Sueiro25}, is another promising research direction.

Topological insulators~\cite{Qi11, Hasan10} are quantum materials with an insulating bulk that can support symmetry-protected gapless edge or surface states, which can only be removed by closing the gap or breaking the relevant symmetry. The tenfold way~\cite{Chiu16, Schnyder08, Kitaev09} classifies free-fermion phases based on time-reversal, charge-conjugation, and chiral symmetries, with distinct phases characterized by topological invariants. 
Beyond this framework, additional symmetries, such as point-group symmetries~\cite{Hughes11, Song17}, can also protect topological phases. While topological invariants and classifications are typically formulated for free-fermion systems, they have been extended in certain cases to interacting systems~\cite{Rachel18, Fidkowski10, Gurarie11, Essin11, Manmana12, Wang12, Tada24, Aligia23, LEHUR2025}.\\

A prototypical model for a topological insulator in one dimension is the Su-Schrieffer-Heeger (SSH) chain~\cite{Su79}. 
This model represents a dimerized chain with alternating hoppings between nearest-neighbor sites. Depending on the relative strength of the two hopping amplitudes, an open SSH chain can feature two mid-gap edge states protected by chiral symmetry and characterized by a winding number~\cite{asboth2016short}. The model also exhibits a quantized Zak phase protected by inversion symmetry. 
Experimentally, it has been realized using graphene nanoribbons~\cite{Franke18, Rizzo18}, ultracold atoms~\cite{Atala13, Meier2016}, and circuits~\cite{Lee18}. In particular, the SSH model has recently become an important playground for exploring the effects of cavity coupling on topological insulators~\cite{Dmytruk22, Shaffer24, Perez25, Ciuti24, Sueiro25}.\\ 
Coupling a topological system with a cavity raises new questions on how to characterize topological protection when photonic operators are involved. 
One possibility is to assume mean-field decoupling between electrons and photons \cite{Dmytruk22}, which allows one to analyze topology in the electron sector using the standard single-particle topology framework. Subsequent works have highlighted the importance of including light-matter correlations and studied topological markers defined on the full light-matter system,  via computing entanglement entropy spectrum of the ground state~\cite{Shaffer24}, a light-matter topological marker that coincides with the ensemble geometric phase~\cite{Ciuti24}, and many-body Green’s function with fermionic and photonic components~\cite{Perez25}. Moreover, breaking of the chiral symmetry due to light-matter correlations was reported in~\cite{Perez25}.\\

In this work, we  investigate how cavity-mediated interactions modify topological properties by studying an effective fermionic Hamiltonian obtained by applying the high-frequency expansion (HFE)~\cite{Mikami16, Sentef20,Li20, Li22, Eckardt15, Sueiro25} to an SSH model coupled to a single-mode cavity. The HFE is valid in the limit of a high-frequency, off-resonant cavity, but is non-perturbative in the light-matter coupling strength, and therefore allows to study strong coupling regimes. It yields an effective, fully electronic Hamiltonian, where hopping amplitudes are renormalized by the cavity, and where cavity-mediated interactions appear in the form of non-local correlations between hopping terms. 
This effective Hamiltonian allows us to analyze to what extend the chiral and inversion symmetries that protect the topological phase of the isolated SSH model are affected by light-matter correlations. We further employ exact diagonalization (ED) on finite-size systems to compute different interacting topological marker: correlation functions between edges on an open chain, a winding number based on the single-particle Green's function~\cite{Gurarie11, Manmana12, Essin11}, and bulk electric polarization via the many-body formula by Resta~\cite{Resta98}. This allows us to establish the topological phase diagram of the effective interacting system.
 The paper is structured as follows. In Sec.~\ref{Model}, we introduce the SSH Hamiltonian, how it is coupled to the cavity, and derive the high-frequency effective Hamiltonian. In Sec.~\ref{EdgeStates}, we study electron-electron correlations on open chains as a marker for edge states. In Sec.~\ref{WindingNumber}, we comment on chiral symmetry in the effective Hamiltonian and compute a winding number based on the single particle Green's function. In Sec.~\ref{Polarization}, we compute bulk polarization as an alternative topological marker. We conclude in Sec.~\ref{Conclusions}.

\section{Model} \label{Model}
\subsection{Review of the SSH model}
For completeness, we start this section by briefly reviewing the SSH model~\cite{Su79,asboth2016short}, before discussing its coupling to light. The SSH model is a one-dimensional tight-binding model with two alternating nearest-neighbor hopping amplitudes. For a chain with $2L$ sites, the Hamiltonian is
\begin{equation}
    H = v \sum_{j=1}^{L} \left(c_{j,A}^\dagger c_{j,B} + \text{h.c.}\right) - w \sum_{j=1}^{L-1}\left(c_{j+1,A}^\dagger c_{j,B} + \text{h.c.}\right),\label{eq:ssh1}
\end{equation}
where $L$ is the number of unit cells, and the two sites within the unit cell are labeled $A$ and $B$. We fix the length of a unit cell to $1$ and the distance between $A$ and $B$ within the unit cell to $0 \leq d <1$~\cite{Platero2013SSH} (see Fig.~\ref{fig:system}). This distance does not enter the Hamiltonian of Eq.~\eqref{eq:ssh1} explicitly and its energy spectrum is independent of $d$. The intra-cell hopping is denoted $v$ and the inter-cell hopping $w$. In the following, we always consider $v,w >0$. \\

Under periodic boundary conditions (i.e., closing the chain with $2L$ sites to a ring), the SSH model can be described by the following periodic Bloch Hamiltonian 

\begin{equation}
    \mathcal{H}(k) = \begin{pmatrix}
        0 & v-we^{-ik} \\
        v-we^{ik} & 0
   \end{pmatrix}.
      \label{eq:ssh2} 
\end{equation}
The form of this Hamiltonian depends on the choice of unit cell~\cite{Fuchs21, Cayssol21}, which we consider fixed by the boundary choice of the open chain in Eq.~\eqref{eq:ssh1} (see also Fig.~\ref{fig:system}).
It has eigenvalues 
\begin{equation}
    E_{\pm}(k)=\pm \sqrt{v^2+ w^2+2vw\cos(k)},
\end{equation}
showing that the bulk gap closes for $w=v$,
and eigenvectors
\begin{equation}
    \ket{\textnormal{v}_{\pm}(k)}=\frac{1}{\sqrt{2}}
    \begin{pmatrix}
        1 \\ \pm e^{i\phi_k}
   \end{pmatrix},
\end{equation}
with $\phi_k = \arg (v-we^{-ik})$.
The SSH model is invariant under chiral (or sublattice) symmetry, which is expressed in momentum space as 
\begin{equation}
    \Gamma: \sigma^z \mathcal{H}(k)\sigma^z = -\mathcal{H}(k) ,\label{eq:cs}
\end{equation}
where $\sigma^z$ is the Pauli $z$ matrix.

Chiral symmetry constrains the Bloch Hamiltonian to describe a closed curve in a plane spanned by $\sigma^x, \sigma^y$ as the wave vector $k$ varies across the Brillouin zone. This allows us to define a winding number, which counts how many times this curve encircles the origin of the plane: 
\begin{equation}
    \nu = \dfrac{1}{2\pi}\int_{-\pi}^\pi
 dk \partial_k \phi_k,
 \label{eq:wind1}
 \end{equation}
 which leads to $\nu=0$ when $v>w$ and $\nu=1$ when $w>v$. This value of the winding number in the bulk can be associated with the number of zero-energy modes localized  at the edges of the chain~\cite{Qi11, Hasan10} described by Eq.~\eqref{eq:ssh1}.

Another quantity that can be computed in one dimension is the Zak phase~\cite{Zak}

\begin{equation}
Z = \int dk \bra{u_{-}(k)}i\partial_k\ket{u_{-}(k)}\, \, \textnormal{mod} \, \,2\pi.
\end{equation} 
It depends continuously on the real-space origin, and is calculated from the cell-periodic Bloch states~\cite{Fuchs21} given by 
\begin{equation}
    \ket{u_{\pm}(k)}= e^{ikx}\ket{\textnormal{v}_{\pm}(k)},
\end{equation} 
where $x$ is the position operator within the unit cell.
The SSH chain has spatial inversion symmetry, which can be expressed in momentum space as 
\begin{equation}
\mathcal{I}:\sigma^x \mathcal{H}(-k) \sigma^x = \mathcal{H}(k).
\end{equation}
The inversion centers are located in the middle of the bonds between $A$ and $B$ either inside a unit cell, or between two unit cells. 
Inversion symmetry leads the Zak phase to be quantized to two possible values mod $2\pi$. 
When the origin is chosen at the inversion center inside the unit cell (i.e., in the $j$th unit cell $x_{j,A} = j-d/2$ and $x_{j,B} = j+d/2$), one finds $Z=0$ when $v>w$ and $Z=\pi$ when $v<w$.
Moreover, the Zak phase is related to the electronic contribution to polarization~\cite{Vanderbilt1}:

\begin{equation}
    P_e= -e\frac{Z}{2\pi},
\end{equation}
where $e$ is the electronic charge, and is hence defined mod $e$. In general, to ensure charge neutrality of the system, the electric polarization should be computed as $P = P_e + P_{\textnormal{ion}}$, i.e., including some positively charged ionic background
~\cite{Fuchs21, Tada24}. This ionic contribution vanishes (mod $e$) if the origin is chosen at the inversion center. The polarization $P$ is either $0$ (for $v>w$) or $e/2$ (for $v<w$) mod $e$~\cite{note}. In the rest of the paper, we set $e=1$.

\subsection{SSH model coupled to a cavity}
\begin{figure}[t]
    \centering
    \includegraphics[scale=0.6]{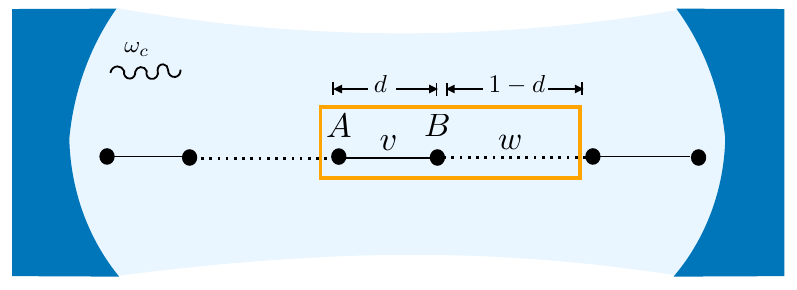}
    \caption{SSH chain placed inside a single-mode cavity with frequency $\omega_c$. The two sites within a unit cell are labeled $A$ and $B$, respectively. The distance between an $A$ and a $B$ site is $d$, and the length of a unit cell is $1$. Our choice of unit cell is indicated by the orange box. For unit cell $j$, the positions of the lattice sites are expressed relative to the origin of the unit cell. Here, $x_{j,A}=j$ and $x_{j,B} = j +d$. For a light-matter coupling strength $g$ and a chain with $L$ unit cells, the intracell hopping $v$ and the intercell hopping $w$ get dressed with a phase as $v\rightarrow ve^{i\frac{gd}{\sqrt{L}}(a+a^{\dagger})}$ and $w\rightarrow we^{i\frac{g(1-d)}{\sqrt{L}}(a+a^{\dagger})}$.  }
    \label{fig:system}
\end{figure}
In the following, we consider an SSH chain with $2L$ sites and a distance $0 \leq d < 1$ between the two sublattices $A$ and $B$~\cite{Platero2013SSH}, coupled to a cavity (see Fig.~\ref{fig:system} for an illustration). The cavity hosts a single mode of frequency $\omega_c$ with a field described by the uniform vector potential $\mathcal{A}= A_0(a+a^{\dagger})$, where $a^{\dagger}$, $a$ are the photonic creation and annihilation operators. The bare cavity Hamiltonian is given by 
\begin{equation}
    H_c = \omega_c\left(a^{\dagger}a + \dfrac{1}{2}\right).
\end{equation}
The coupling to the cavity is introduced via Peierls substitution~\cite{dmytruk2021gauge, Dmytruk22, Perez25, Ciuti24, Shaffer24, Bacciconi24, Sueiro25}, which dresses the hopping amplitudes with a phase proportional to both the vector potential and the distance between neighboring sites. This leads to the light–matter Hamiltonian~\cite{Dmytruk22, Perez25, Ciuti24, Shaffer24, Sueiro25}
 \begin{align} \label{eq:Fullham}
     H_{\textnormal{LM}} &= \omega_c \left(a^\dagger a + \dfrac{1}{2}\right) + v \sum_{j=1}^L \left(e^{i g_1 (a+a^{\dagger})}c_{j,A}^\dagger c_{j,B} + \text{h.c.}\right) \nonumber\\&-w \sum_{j=1}^{L-1}\left(e^{-i g_2(a+a^\dagger)}c_{j+1,A}^\dagger c_{j,B} + \text{h.c.}\right),
 \end{align}
 where $g_1 = g d/\sqrt{L}$ and  $g_2 = g(1- d)/\sqrt{L}$.
Through the Peierls phase, this Hamiltonian now depends explicitly on $d$, contrary to the uncoupled case, and the two hopping terms are renormalized differently unless $d=1/2$.
The scaling of the coupling strength $g=A_0$ with the system size $L$ is necessary for a well-defined thermodynamic limit \cite{Pilar20} in the strong coupling regime. When $L\rightarrow \infty$, the single cavity mode does not affect the bulk properties of the system. Nevertheless, the cavity can have strong effects on the finite-length properties \cite{Dmytruk22, Sueiro25, Ciuti24}.\\


\subsection{High-frequency expansion}

In order to derive an effective, electron-only Hamiltonian, we follow Ref.~\cite{Sueiro25} and apply the high-frequency expansion (HFE)~\cite{Sentef20, Li20, Li22,fernandez25, Eckardt15, Mikami16} to Eq.~\eqref{eq:Fullham}, valid when the cavity frequency $\omega_c$ is the largest energy scale in the system. 
The high-frequency expansion yields an effective fermionic Hamiltonian acting within photon-number subspaces $\ket{n}$ ($n$ ranges from 0 to infinity) which reads

\begin{align} \label{eq:heff1}
H^{\textnormal{eff, n}}&=\bra{n} H_{\textnormal{LM}} \ket{n}\!\\ \nonumber
&+\! \dfrac{1}{\omega_c}\!\sum_{l\neq n}\dfrac{\bra{n}H_{\textnormal{LM}}\ket{l}\bra{l}H_{\textnormal{LM}} \ket{n}}{(n-l)} +\!\mathcal{O}\!\!\left (\frac{1}{\omega_c^2}\right).
 \end{align}
Evaluating the matrix elements of the Peierls phase yields \cite{Li20, Li22} 
\begin{align}\label{eq:hnl}
  \bra{n} H_{LM}\ket{l} =&\omega_c \left(n+\frac{1}{2}\right) \delta_{n,l}I_{el}\\&+v j_{n,l}\nonumber(g_1)\sum_j (i^{\vert n-l \vert} c_{j,A}^{\dagger}c_{j,B} + \text{h.c.})\\& \nonumber + wj_{n,l}(g_2)\sum_j (i^{\vert n-l \vert}c_{j,B}^{\dagger}c_{j+1,A} + \text{h.c.}),
\end{align} 
where
\begin{equation}
  j_{n,l}(x) = e^{-x^2/2}\!\!\!\!\sum_{k = 0}^{\min{(n, l)}} \!\!\!\!\!\!\frac{(-1)^k( x )^{2k+\vert n-l\vert}\sqrt{l! n!}}{k! (k+\vert n-l \vert)!(\min{(n, l)}-k)!}
    \label{eq:alpha}
\end{equation}
and $I_{el}$ is the identity over the electronic Hilbert space.
 The ground state of the system lies in the $n=0$ photon subspace~\cite{Li22, Sueiro25}. Thus, in the following, we concentrate on this subspace and denote $ H^{\textnormal{eff}, 0}\equiv  H^{\textnormal{eff}}$, leading to the effective Hamiltonian 
\begin{widetext}
\begin{align}\label{eq:Heff1}&H^{\textnormal{eff}} =\frac{\omega_c}{2}\!+\!v_{\textnormal{eff}}\!\sum_{j=1}^L (c_{j,A}^\dagger c_{j,B} \!+\!\textnormal{h.c.}\!)\!-\!w_{\textnormal{eff}}\!\sum_{j=1}^{L-1}(c_{j+1,A}^\dagger c_{j,B}\!+\!\textnormal{h.c.}\!)\\\nonumber+&\frac{1}{\omega_c}\sum_{m,j}v_{\textnormal{eff}}^2 f(g_1^2)(c_{j,A}^\dagger c_{j,B} c_{m,A}^\dagger c_{m,B}  +c_{j,B}^\dagger c_{j,A} c_{m,B}^\dagger c_{m,A}) +  v_{\textnormal{eff}}^2 f(-g_1^2)(c_{j,A}^\dagger c_{j,B} c_{m,B}^\dagger c_{m,A}  +c_{j,B}^\dagger c_{j,A} c_{m,A}^\dagger c_{m,B}) \\ \nonumber+\!& w_{\textnormal{eff}}^2f(\!g_2^2)(c_{j\!+\!1,A}^\dagger c_{j,B} c_{m\!+\!1,A}^\dagger c_{m,B}\!  +\!c_{j,B}^\dagger c_{j\!+\!1,A} c_{m,B}^\dagger c_{m\!+\!1,A}) \!+\!  w_{\textnormal{eff}}^2 f(\!-\!g_2^2)(c_{j\!+\!1,A}^\dagger c_{j,B} c_{m,B}^\dagger c_{m\!+\!1,A} \!+\!c_{j,B}^\dagger c_{j\!+\!1,A} c_{m\!+\!1,A}^\dagger c_{m,B}) \nonumber \\\nonumber
-&v_{\textnormal{eff}}w_{\textnormal{eff}}f(g_1g_2)(c_{j,A}^{\dagger}c_{j,B}c_{m,B}^{\dagger}c_{m+1,A}+c_{j,B}^{\dagger}c_{j,A}c_{m+1,A}^{\dagger}c_{m, B}+c_{j+1,A}^{\dagger}c_{j,B}c_{m,B}^{\dagger}c_{m,A}+c_{j,B}^{\dagger}c_{j+1,A}c_{m,A}^{\dagger}c_{m,B})\nonumber \\\nonumber
-&v_{\textnormal{eff}}w_{\textnormal{eff}}f(-g_1g_2)(c_{j,A}^{\dagger}c_{j,B}c_{m+1,A}^{\dagger}c_{m,B}+c_{j,B}^{\dagger}c_{j,A}c_{m,B}^{\dagger}c_{m+1,A}+c_{j+1,A}^{\dagger}c_{j,B}c_{m,A}^{\dagger}c_{m,B}+c_{j,B}^{\dagger}c_{j+1,A}c_{m,B}^{\dagger}c_{m,A}),\end{align}
\end{widetext} 

 where $v_{\textnormal{eff}}=v e^{-g_1^2/2}$ and $w_{\textnormal{eff}}=w e^{-g_2^2/2}$, and 
 \begin{equation} 
 f(x)=\sum_{l>0}\frac{(-1)^{l+1}}{l (l!)}x^l.
\end{equation} 
At zeroth order in the expansion, the effective Hamiltonian is simply an SSH Hamiltonian with renormalized hoppings $v_{\textnormal{eff}}$ and $w_{\textnormal{eff}}$, and the effect of the cavity simply amounts to shifting the critical ratio of hopping amplitudes from $w/v=1$ to $w/v = e^{\frac{g^2}{2L}(1-2d)}$~\cite{Sueiro25}. However, the first order contribution, in $1/\omega_c$, contains long-range correlations between hopping terms.
Due to these interaction terms, $H^{\textnormal{eff}}$ cannot be analyzed within a single-particle framework unless further approximations are made. In the following, we will consider finite-size SSH chain and compute many-body topological marker for $H^{\textnormal{eff}}$ including these interactions using ED. This differs from the approach taken in Ref. \cite{Sueiro25}, where the effective HFE Hamiltonian is derived in momentum space and interaction terms are treated within Hartree-Fock mean-field (see also later in Fig.~\ref{fig:polwindswitch} for a comparison between the two approaches).

\section{Edge states}\label{EdgeStates}
\begin{figure}
    \centering
\includegraphics[width=\linewidth]{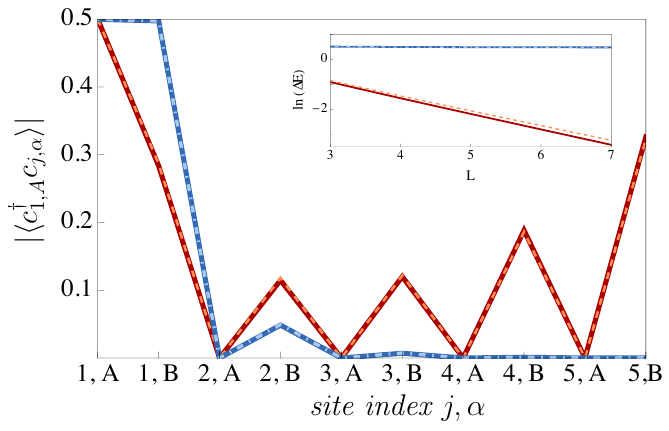}
\caption{Comparison of two-point correlation function $|\langle c_{1,A}^\dag c_{j,\alpha} \rangle|$, with $j=1,...,L$ and $\alpha = A,B$, computed on the full light-matter ground-state (LM)  and the HFE groundstate (HFE) as a function of the lattice site $j,\alpha$ for two values of $w/v$. 
   Solid red and solid dark blue lines were obtained using HFE for $w/v = 1.8$ and $w/v=0.2$, respectively. 
   Dashed orange and dotdashed lightblue lines were computed on the LM ground state for $w/v = 1.8$ and $w/v=0.2$, respectively.
   For both choices of hopping amplitude $w/v$, the correlation function computed for the HFE ground state and the one computed for the LM ground state show very good agreement.
   The correlation is computed between edge site $1$ and the other sites of a chain with $2L=10$ sites. 
    In the inset, we show the energy gap (in log scale) between the two lowest energy states, computed at half-filling, as a function of the number of unit cells $L$.
    Other parameters are fixed as the intra-cell distance $d=0.5$, cavity frequency $\omega_c/v=10$, light-matter coupling $g=1$, and photonic cutoff $N_c=5$ \cite{note2}.
    }
    \label{fig:compare}
\end{figure}
An uncoupled SSH chain with $2L$ sites under open boundary conditions hosts two edge states whose energy goes to zero when $L \rightarrow \infty$, when the intercell hopping $w$ is larger than the intracell hopping $v$~\cite{asboth2016short}. In order to verify the presence of edge states for the interacting Hamiltonian $H^{\textnormal{eff}}$, we compute the two-point correlation~\cite{Defossez25, Shaffer24, Bacciconi24} $\bra{\psi_0^{\textnormal{eff}}}c_{1,A}^{\dagger}c_{j, \alpha}\ket{\psi_0^{\textnormal{eff}}}$ between the left edge of the chain (site $A$ in unit cell $1$) and a site $j, \alpha$, with $1 \leq j\leq L$ and $\alpha \in \lbrace A, B \rbrace $ (see Fig.~\ref{fig:compare}), in the presence of a non-zero coupling to the cavity ($g=1$). The expectation value is taken on the lowest-energy eigenstate of $H^{\textnormal{eff}}$, computed at half-filling. When $v \ll w$, we find that the correlation function decays rapidly to zero when computed for $j$ distant from the site $1, A$ on the edge of the chain. The value of $1/2$ when $j, \alpha = 1,A$ is due to the ground state being at half-filling. On the other hand, when $w\gg v$, the correlation function oscillates in the bulk and increases again as $j$ gets closer to the opposite edge of the chain, showing non-local correlations between edges. In this case, we find a second low-energy state, whose splitting with the lowest energy state decreases exponentially when the system size increases (see inset in Fig.~\ref{fig:compare}). This is in agreement with the presence of two edge states in the topological phase of the SSH model: in the ground state, at half filling, only one of them is occupied, leading to a ground-state degeneracy of two in the thermodynamic limit. For a centrosymmetric chain, both edge states are half filled, and the two ground states correspond to a symmetric and antisymmetric superposition of the occupied and empty state. 

In Fig.~\ref{fig:compare}, we also display the results for the two-point correlation function $\bra{\psi_0^{\textnormal{LM}}}c_{1, A}^{\dagger}c_{j, \alpha}\ket{\psi_0^{\textnormal{LM}}}$ computed on the ground state of the full light-matter Hamiltonian~\eqref{eq:Fullham}. For cavity frequencies \textcolor{red}{$\omega_c/v \geq 5$}, these results are in very good agreement with the ones computed from the ground state of the HFE Hamiltonian, which confirms the validity of the high-frequency expansion to study the cavity-coupled system.\\
In the following two sections, we switch from the open-chain geometry to a ring with periodic boundary conditions, in order to compute topological markers for the SSH chain in the bulk.

\section{Winding number}\label{WindingNumber}
\begin{figure}[t]
    \centering
    \includegraphics[width=\linewidth]{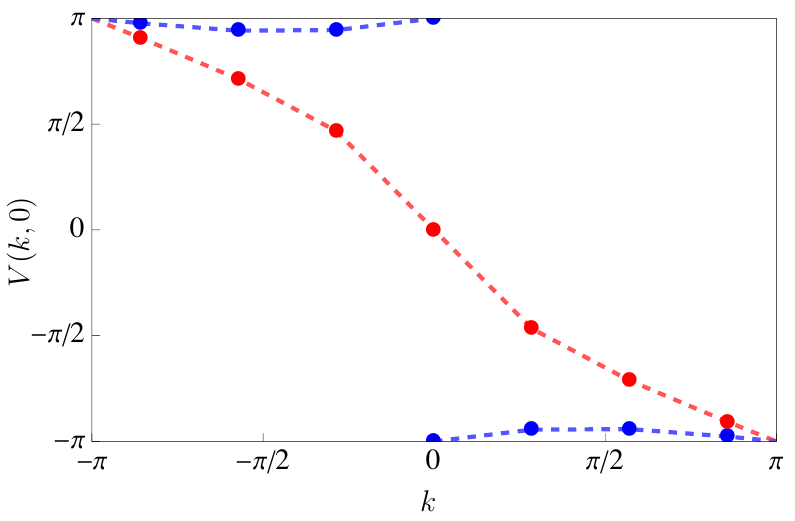}
      \caption{$V(k,0)$ plotted as a function of $k$ between $-\pi$ and $\pi$ for an SSH chain with $L = 7$ unit cells and sublattice spacing $d=0.5$, coupled to a cavity with frequency $\omega_c/v=10$, coupling strength $g=1$ and photonic cut-off $N_c = 30$. The blue dots correspond to the values taken by $V(k,0)$ for $w/v =0.2$ and the red dots correspond to the values taken by $V(k,0)$ for $w/v=1.8$.}
    \label{fig:winding}
\end{figure}

For a free fermion system, the presence of chiral symmetry is associated with the existence of a quantized winding number that can be computed in the bulk from the Bloch Hamiltonian (see Eq.~\eqref{eq:ssh2}).

For an interacting system, although Bloch’s theorem no longer applies, an analogous relation to the chiral symmetry condition satisfied by the Bloch Hamiltonian (see Eq.~\eqref{eq:cs}) can be established between a unitary matrix and the imaginary-time single-particle Green’s function  $G(k,\Omega)$~\cite{Gurarie11, Manmana12}:
\begin{equation}
    \Gamma G(k,\Omega) \Gamma = - G(k, -\Omega),
\label{eq:GFsym}
\end{equation}
where $k$ is the wave vector and $\Omega$ the imaginary frequency. In the presence of such a symmetry, a quantized topological invariant can be defined as~\cite{Volovik, Gurarie11, Essin11, Manmana12, Jin24, Pavan23}
\begin{equation}
\mathcal{\nu} = \dfrac{1}{4\pi i}\int_{-\pi}^{\pi} dk  \textnormal{Tr} \left ( \Gamma G^{-1}(0, k) \partial_k G(0,k) \right ),
\label{eq:gf}
\end{equation}
where the Green's function is evaluated at zero frequency. In fact, this expression reduces to the winding number Eq.~\eqref{eq:wind1} in the non-interacting case, when  $G(\Omega,k)=(\Omega \mathbf{1} - \mathcal{H}(k))^{-1}$.  
In the rest of this section, we show that a generalized version of chiral symmetry holds for the HFE Hamiltonian and that, in consequence, we can compute a well-defined winding number using the single-particle Green's function. 

\subsection{Chiral symmetry and HFE Hamiltonian}
In order to analyze the symmetry of the HFE Hamiltonian Eq. \eqref{eq:Heff1}, we use the fact that the unitary chiral symmetry operator $\Gamma$ can also be expressed in real space through its action on the second quantized Hamiltonian $H$~\cite{Zirnbauer21, Chiu16, Gurarie11}:
\begin{equation}
     H = \Gamma^{\dagger}H^* \Gamma,
     \label{eq:comm}
\end{equation}
where $H^*$ is the complex conjugate of $H$ and $\Gamma$ acts on the fermionic creation and annihilation operators as  
\begin{equation}
\Gamma^{\dagger}c_{j, A} \Gamma =c_{j,A}^{\dagger}, \, \, \,\Gamma^{\dagger}c_{j, B} \Gamma =-c_{j, B}^{\dagger}. \label{eq:symm2}
\end{equation}
Here, the chiral symmetry operator $\Gamma$ exchanges creation and annihilation operators, similar to a particle-hole symmetry~\cite{Zirnbauer21}. For the isolated SSH model, this reduces to the standard anticommutation relation $\Gamma^{\dagger} h\Gamma = -h$ on the first-quantized Hamiltonian $h$~\cite{Gurarie11, Chiu16, Zirnbauer21}. As demonstrated in Ref.~\cite{Gurarie11}, Eqs.~\eqref{eq:comm} and~\eqref{eq:symm2} lead to the symmetry requirement of Eq.~\eqref{eq:GFsym} in momentum space.\\
With the above definition, it can easily be verified using Eq. \eqref{eq:hnl} that
\begin{equation}
    \Gamma^{\dagger} \bra{n}H_{\textnormal{LM}}\ket{l}\Gamma = \bra{n}H_{\textnormal{LM}}\ket{l}^* \, \, \, \forall n,l,
\end{equation} 
from which it follows directly that the zero-order term in $H^{\textnormal{eff}}$ is invariant under chiral symmetry. The first-order term involves a product of terms  $\bra{n} H_{LM} \ket{l}$. Since each of these terms individually verifies chiral symmetry, using unitarity of $\Gamma$ and $(AB)^* = A^*B^*$, one can show that their product verifies chiral symmetry too. Higher order terms of the high-frequency expansion similarly involve products of terms $\bra{n} H_{LM} \ket{l}$ (see Ref. \cite{Mikami16, Sueiro25} for details on the HFE), indicating that chiral symmetry is not broken by the coupling to the cavity in the high-frequency limit to all orders in the expansion as long as this structure is conserved.\\ 
The chiral symmetry defined in Eq. \eqref{eq:cs} is a property of the many-body Hamiltonian acting in $2^{2L}$-dimensional electronic Fock space, and  relates the $N$-particle Hilbert space to the $2L-N$ particle Hilbert space. It guarantees the existence of a well-defined winding number computed for the many-body ground state at half-filling \cite{Gurarie11}, which we compute below. However, chiral symmetry might appear broken when restricting the study to a Hilbert space with arbitrary particle number $N$. This is consistent with results of Ref. \cite{Perez25}, which observes that chiral symmetry is broken  through coupling to the cavity within the single-particle subspace \cite{note3}. 
\subsection{Winding number for HFE Hamiltonian}
As a consequence of Eq.~\eqref{eq:GFsym} and $\Gamma= \sigma_z$ in momentum space, the Green's function at zero frequency takes the following form~\cite{Gurarie11}

\begin{equation}
 G(k,0)=   \begin{pmatrix}
        0 & W(k, 0) \\ W(k,0)^* & 0 
    \end{pmatrix}.
\end{equation}
For a finite size system with $L$ unit cells, we compute the off-diagonal component $W$ as 
\begin{equation}
    W(k,0) = \dfrac{1}{L}\sum_{i, j}e^{-ik(i-j)}G_{A,B;i,j}(0),
\end{equation}
which is the discrete Fourier transform for a finite system with $L$ unit cells, where $k=\frac{2\pi n}{L}$, with $n=0, 1, ..., L-1$, and $G_{A,B; i,j}(\Omega) = \int_0^{\infty} d\tau e^{i\Omega \tau}\langle c_{i,A}(\tau)c_{j,B}(0)\rangle$ is computed from its spectral decomposition (see Ref.~\cite{Gurarie11} for more details). Note that here we keep the same unit cell convention as for the Bloch Hamiltonian of Eq.~\eqref{eq:ssh2}. \\
The winding number in Eq.~\eqref{eq:gf} is then determined by the complex phase of $W(k,0)$~\cite{Manmana12}. Thus, instead of computing $\nu$, we directly compute

\begin{equation}
    V(k, 0) =  \arg\left( W(k,0)\right).
\label{eq:wind}
\end{equation}
In Fig.~\ref{fig:winding} we represent $V(k,0)$ for two different ratios of $w/v$, showing that we indeed find two different phases distinguished by $\nu=0$ and $\nu=1$. When $w/v=0.2$, $V(k,0)$ oscillates in a small range around $\pi$, indicating a winding number of $0$. When $w/v=1.8$, $V(k,0)$ covers the entire interval from $-\pi$ to $\pi$ as $k$ varies from $0$ to $2\pi$, corresponding to a winding number of $1$. 
The winding of $1$ is in agreement with the observation of the previous section of the existence of one occupied edge state in the ground-state ( see also Appendix~\ref{Appa} for more details.)

\begin{figure*}[t]
    \centering
\includegraphics[width=0.9\textwidth]{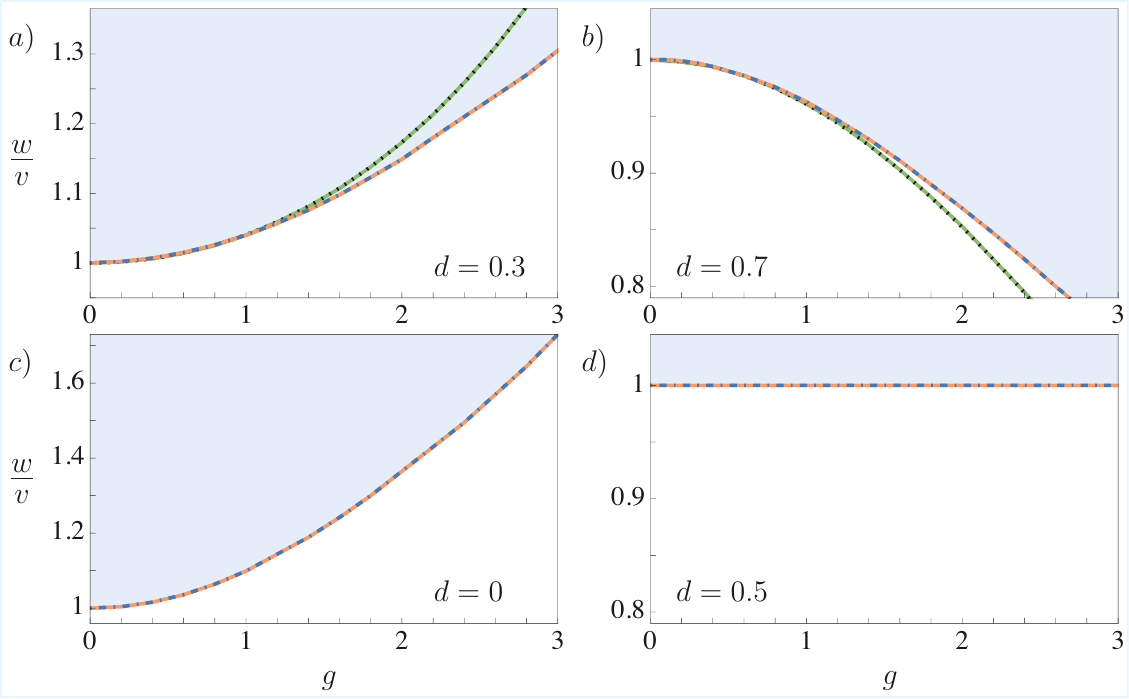}
    \caption{
    Phase diagram for a cavity embedded SSH chain with periodic boundary conditions  as a function of coupling strength $g$ and hopping amplitudes $w/v$ for different values of intra-unit cell distance $d$ (that are indicated in the figure: : $a$) $d=0.3$, $b$) $d=0.7$, $c$) $d=0$, and $d$) $d=0.5$). The light blue area indicates the topological phase for which $\nu=1$ and $P=0.5$. The white area indicates the trivial phase, for which $\nu=0$ and $P=0$. The solid orange lines were obtained from the winding number calculation (Eq.~\eqref{eq:wind}) using ED and mark the transition from $\nu=1$ to $\nu=0$. The dotdashed blue lines were obtained from the calculation of the polarization given by Eq.~\eqref{eq:pol} using ED and mark the transition from $P=0.5$ to $P=0$ (mod 1). In a) and b), the the dotted black line indicates the gap closing point as obtained from the zeroth order of $H^{\textnormal{eff}}$ and the solid green line indicates the gap closing point obtained from $H^{\textnormal{eff}}$ where the interaction term was treated in Hartree-Fock mean-field in momentum space (similar to Ref. \cite{Sueiro25}).  Other parameters are chosen as $L=5$, $\omega_c/v=10$ and $N_c = 30$.
    }
    \label{fig:polwindswitch}
\end{figure*}

Computing the winding number using Eq. \eqref{eq:wind}, 
we arrive at the phase diagrams presented in Fig.~\ref{fig:polwindswitch}, which show that the critical ratio of hopping amplitudes $w/v$ depends on the coupling strength to the cavity $g$ as well as the geometry of the SSH chain through the sublattice distance $d$. In particular, when $d=0.5$, the phase transition point is not affected by the cavity coupling (note that this can already be inferred from the light–matter Hamiltonian Eq. 
\eqref{eq:Fullham}: for $d=0.5$, the Peierls phase factorize, such that the Hamiltonian decomposes as a tensor product with the SSH Hamiltonian, and the gap closing point is  determined by the SSH Hamiltonian Eq. 
\eqref{eq:ssh1}). When $d=0$, the SSH model is equivalent to the chiral Shockley model~\cite{Pershoguba12, Fuchs21}. As shown in Fig.~\ref{fig:polwindswitch}(a) and (b), including the first-order contribution to the HFE Hamiltonian modifies the phase diagram compared to the case where only the zeroth-order contribution is considered, particularly as 
$g$ increases. For the small system sizes studied here, a Hartree–Fock treatment of the interactions (cf. Ref.~\cite{Sueiro25}) yields results that coincide with those obtained from the zeroth-order HFE Hamiltonian.
Note that while our results are obtained for small system sizes, it was shown in Ref.~\cite{Manmana12} that finite-size effects do not significantly impact the calculation of the winding number. \\
In Appendix \ref{app:dep}, we additionally show the phase diagram as a function of cavity frequency and system size. \\
\section{Polarization}\label{Polarization}

In this section, we compute the electric polarization as an alternative topological marker. For a free fermion system, the electric contribution to the bulk polarization can be expressed as the Zak phase~\cite{Vanderbilt1}. The electric polarization can also be computed in the presence of interactions via the real-space formula introduced by Resta~\cite{Resta98}, that remains quantized to two values if inversion symmetry is conserved~\cite{Tada24}. Resta's formula has been proposed as a topological marker for finite and zero temperature topological insulators with and without interactions~\cite{Bardyn18,Moligni23, Huang25, Hannukainen25,Gilardoni22} (where it is also called ensemble geometric phase) and an extension to light-matter systems has been considered in Ref.~\cite{Ciuti24} for a one-dimensional topological insulator coupled to a cavity.

Here, we use this formula to compute the polarization of the HFE Hamiltonian with finite number $L$ of unit cells and periodic boundary conditions. \\ 
The polarization for the charge neutral system is computed as~\cite{Resta98, Tada24, Resta20, Aligia23, Aligia99} 
\begin{equation} 
P = \dfrac{1}{2\pi}\textnormal{Im} \ln \bra{\psi_0}e^{\frac{2 i\pi}{L} \hat{X}}\ket{\psi_0},
\label{eq:pol}
\end{equation}
where  
\begin{equation}
\hat{X}= \sum_{j=1}^L x_{j,A}\left(c_{j,A}^{\dagger}c_{j,A}-\dfrac{1}{2}\right)+x_{j,B}\left(c_{j,B}^{\dagger}c_{j,B} - \dfrac{1}{2}\right).
\label{eq:X}
\end{equation} 
Here, $x_{j,A}$ and $x_{j,B}=x_{j,A}+d$ denote the positions of the sublattice sites $A$ and $B$ within the unit cell $j$ (their exact expression depends on the choice of real-space origin) and the constant $-1/2$ shifts arise from including a uniform ionic background to ensure charge neutrality. In other words, $\hat{X}$ represents the position of the electronic charge distribution relative to a uniform positive ionic background. Including the background contribution also makes the value of $P$ independent of the choice of real-space origin \cite{Watanabe18}. The expectation value in Eq.~\eqref{eq:pol} is taken on the many-body fermionic ground state. \\
In the presence of inversion symmetry, $P$ maps to $-P$ (see, e.g.,~\cite{Fuchs21, Tada24} 
); as $P$ is defined mod $1$ via Eq.~\eqref{eq:pol}, this means that in the presence of inversion symmetry there are only two possible branches for the polarization, $0$ mod $1$ or $1/2$ mod $1$. This quantization appears for any system size under periodic boundary conditions~\cite{Resta20}. 
\\For a chain with $L$ unit cells labeled $1, 2, ..., L$, inversion about the center of the chain acts on the fermionic operators as 
\begin{equation}
    \mathcal{I}c_{j,A/B} \mathcal{I}^{-1} =  c_{ L+1-j, B/A}.
    \label{eq:inversion}
\end{equation}
We verified that, for both  even and odd $L$, $\mathcal{I}$ commutes with the Hamiltonian Eq. \eqref{eq:Heff1}.
\\ 
Computing polarization via Eq.~\eqref{eq:pol} using ED of the HFE Hamiltonian~\eqref{eq:Heff1}, we find that the values of $w/v$ for which $P$ changes from $1/2$ to $0$ is in perfect agreement within numeric precision (up to $10^{-3}$) with the values for which the winding number changes from $1$ to $0$ (see Fig.~\ref{fig:polwindswitch}).
The existence of a quantized polarization is related to the presence of inversion symmetry and does not guarantee by itself the existence of gapless edge states~\cite{Tada24, Song17}. However, in the presence of chiral symmetry, polarization can be used as an alternative topological marker indicating the presence of gapless edge states. \\

\section{Conclusions}~\label{Conclusions}
In this work, we have studied an SSH chain coupled to an off-resonant single mode cavity by using the high-frequency expansion and computing different topological markers for an effective interacting fermionic Hamiltonian, thus allowing us to use the existing framework of interacting topological phases to investigate topological properties. We find that chiral symmetry holds for the effective Hamiltonian, allowing one to compute a well-defined and quantized winding number in the bulk. Despite the long-range interactions generated by the cavity, we observe that the winding number remains quantized to either $0$ or $1$. By computing electron-electron correlation functions between the edges of an open chain, we further verify that the existence of a non-trivial winding number still corresponds to the presence of edge states. 
Finally, we compute the polarization under periodic boundary conditions, which is quantized to two values due to spatial inversion symmetry of the effective Hamiltonian. The phase diagram obtained from polarization coincides exactly with that obtained from the winding number, similarly to the case of the SSH model without cavity coupling. However, through cavity coupling, the point of phase transition is shifted depending on the intradimer distance $d$, the coupling strength $g$, the cavity frequency $\omega_c$, and the size $L$ of the chain, in agreement with the results obtained in Refs. \cite{Dmytruk22, Ciuti24, Sueiro25}. \\
Our work shows that the cavity-mediated many-body interactions are important for conserving the protecting symmetries of the original phase, and highlight the importance of using a many-body approach for studying the groundstate of the light-matter system. 

\section*{Acknowledgements}
We thank Alvaro Gómez-León, Jean-Noël Fuchs, Marco Schirò, Sarath Prem and Jules Sueiro for helpful discussions. This work is supported by ERC grant Q-Light-Topo (Grant Agreement No. 101116525).\\
\appendix

\section{Local edge spectral functions}\label{Appa}
\begin{figure}[t]
    \centering
\includegraphics[width=0.49\textwidth]{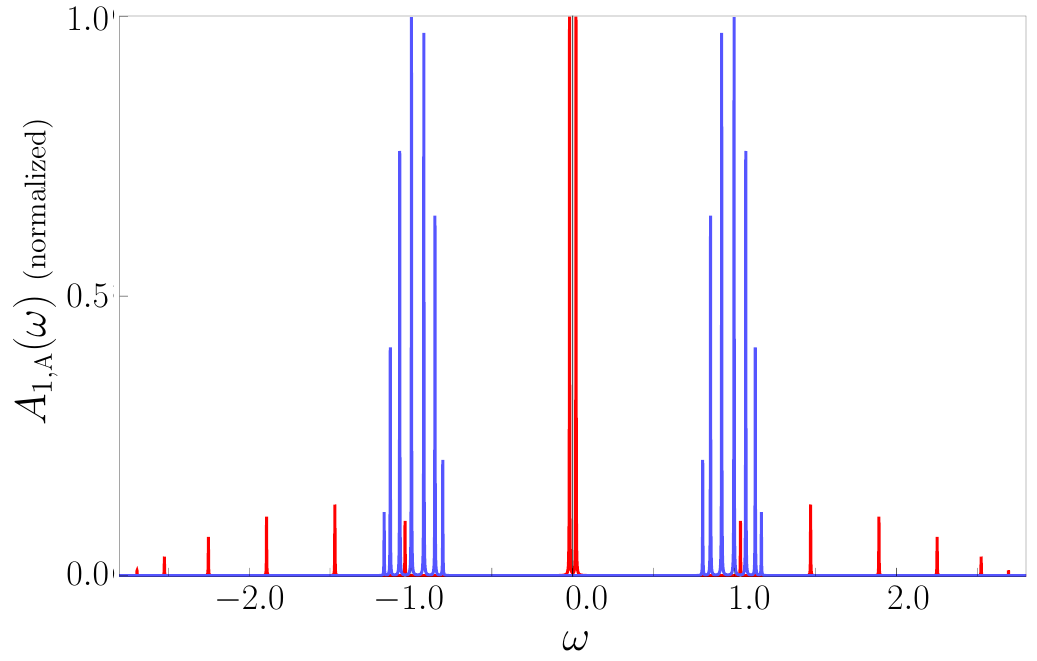}
    \caption{Local edge spectral function $A_{1,\textnormal{A}}(\omega)$ for the left edge (site $1,$A) of an interacting SSH chain of length $L=7$, computed from the HFE Hamiltonian Eq. \eqref{eq:Fullham} using ED, with $g=1$, $d=0.5$, $N_c = 30$, and $\omega_c/v=10$, as a function of real frequency $\omega$. In blue, $A_{1,\textnormal{A}}(\omega)$ for $w/v =0.2$ (trivial phase) and in red for $w/v =1.8$ (topological phase). $\eta=10^{-4}$. The spectral function is normalized by its maximum value.}
    
    \label{fig:lesf}
\end{figure}
For a non-interacting system the winding number is in direct correspondence with the number of zero-energy edge states per edge \cite{Qi11, Hasan10}. For an interacting system, the winding number can also count edge zeros of the Green's function in addition to edge poles~\cite{Gurarie11, Manmana12, Yoshida14, Wagner23}. In order to verify that the winding number calculated in Sec. IV B corresponds to an edge state, we compute the local edge spectral function $A_{1,\textnormal{A}}=-\frac{1}{\pi}\textnormal{Im} (G_{\textnormal{A},\textnormal{A};1,1}(\omega+i\eta))$ for the same choices of parameters as for the winding numbers of Fig.~\ref{fig:winding}. The results are shown in Fig.~\ref{fig:lesf}. For the choice of parameters where $\nu=1$, we observe the presence of two energy poles of the Green's function very close to zero frequency. The small splitting (of the order of $10^{-2}$) can be understood as a finite size effect, the edge states hybridizing and moving away from exactly zero energy for a finite size system. For an infinite system, we expect the presence of a single pole at zero energy. This is consistent with the presence of an edge state at each edge of the chain and a winding number of $1$. In the trivial phase, the local edge spectral function does not show poles close to zero frequency (see Fig. \ref{fig:lesf}).

\section{Dependence on system size and on cavity frequency}
\label{app:dep}
\begin{figure}[h]
  \centering
  \includegraphics[width=0.45\textwidth]{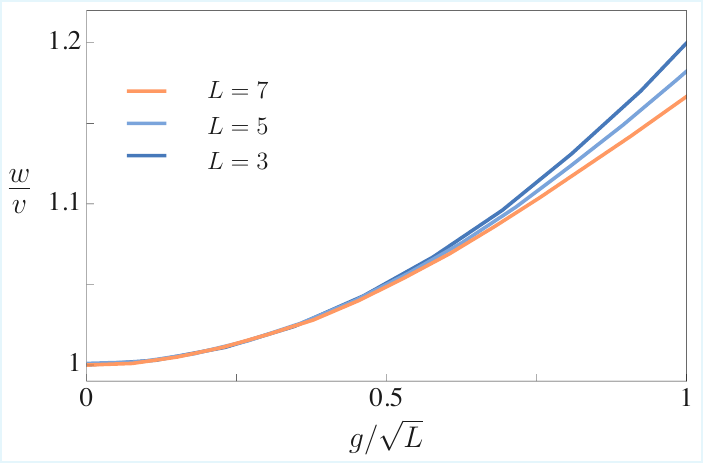}
  \caption{Phase diagram for a cavity embedded SSH chain with periodic boundary conditions obtained from polarization calculation using ED for different system sizes (number of unit cells) $L$, with intra-unit cell distance $d=0.3$, cavity frequency $\omega_c/v=10$ and photonic cutoff $N_c = 30$, as a function of the rescaled coupling strength $g$.}
   \label{fig:app1}
  \end{figure}

\begin{figure}[h]
  \includegraphics[width=0.45\textwidth]{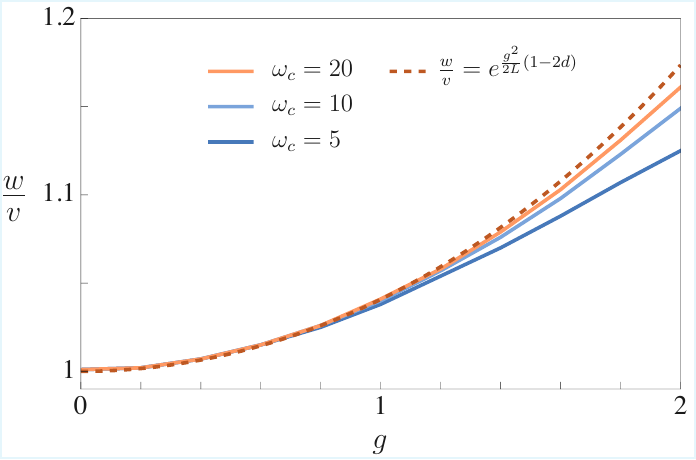}
  \caption{Phase diagram for a cavity embedded SSH chain with periodic boundary conditions obtained from polarization calculation using ED for different cavity frequencies $\omega_c/v$. The number of unit cells is fixed to $L=5$, the intra-unit cell distance $d=0.3$, and photonic cutoff $N_c = 30$. The red dashed line shows the condition of gap closing ($v_{\textnormal{eff}}=w_{\textnormal{eff}}$) for the zeroth order of $H_{\textnormal{eff}}$ which is independent of $\omega_c$.}
  \label{fig:wdep}
\end{figure}

In this appendix, we study the dependence of the phase diagram of the cavity embedded SSH chain with periodic boundary conditions on system size $L$ (see Fig.~\ref{fig:app1}) and on cavity frequency $\omega_c$ (see Fig.~\ref{fig:wdep}). Fig.~\ref{fig:app1} shows that the phase diagram depends on the system size $L$ even when plotted as a function of the variable $g/\sqrt{L}$. This dependence arises from the first order term in $H^{\textnormal{eff}}$, as the zeroth order term scales with $g/\sqrt{L}$.  In Fig.~\ref{fig:wdep}, we compare the boundary between $P=1/2$ and $P=0$ in the $w/v$, $g$ plane as a function of the cavity frequency $\omega_c$, for a fixed system size. As expected, for a large frequency, the behavior of the HFE Hamiltonian is largely dominated by the zeroth order term. For lower frequency and the system sizes considered here, the first order term contribution tends to lower the shift in the topological phase transition induced by the zero order term. These results agree qualitatively with those obtained in Refs.~\cite{Dmytruk22,Sueiro25}. 


%


\begin{thebibliography}{87}
\bibitem [{\citenamefont {Typically}()}]{note}%
  \BibitemOpen
  \bibfield  {author} {\bibinfo {author} {\bibnamefont {Typically}},\
  }\href@noop {} {\bibinfo {title} {the ionic background can be chosen as
  having a $+1/2$ charged ion located at each lattice site \cite{Fuchs21}. {I}n
  this case, {$P$} can change by $1/2$ if one shifts the lattice by half a unit
  cell \cite{Fuchs21}. {H}ere, we consider that our unit cell choice is fixed
  by the open boundary conditions and half-unit-cell translations are not
  allowed.}}\BibitemShut {Stop}%
\bibitem [{not()}]{note2}%
  \BibitemOpen
  \href@noop {} {\bibinfo {title} {The full light-matter {H}ilbert space grows
  as $2^{2L}\times ({N}_c +1)$ where the first factor acocunts for the
  electronic {F}ock space dimension (by fixing the electron occupation to half
  filling, it reduces to $\begin{pmatrix} 2{L} \\{L} \end{pmatrix}$) and the
  second factor to the photonic {F}ock space dimension. {F}or ${L}=3,4,5$, we
  verified that the difference in the electron-electron correlations computed
  with ${N}_c=5$ and a larger ${N}_c=10$ are maximally of order of
  $10^{-10}$.}}\BibitemShut {Stop}%
\bibitem [{\citenamefont {{B}y normal ordering the first-order contributions in
  the {HFE} {H}amiltonian of~{E}q. 16}()}]{note3}%
  \BibitemOpen
  \bibfield  {author} {\bibinfo {author} {\bibnamefont {{B}y normal ordering
  the first-order contributions in the {HFE} {H}amiltonian of~{E}q. 16}},\
  }\href@noop {} {\bibinfo {title} {one can show that they separate into
  one-body terms consisting of on-site energy terms and next-nearest-neighbor
  hopping terms, and actual two-body interaction terms. {W}hen acting on the
  single-particle {H}ilbert space, the two-body interaction terms are zero,
  while the next-nearest-neighbor hopping terms break chiral symmetry by
  coupling sites from the same sublattice.}}\BibitemShut {Stop}%
\end{thebibliography}
\end{document}